\documentclass[aps,prl,showpacs,twocolumn,superscriptaddress]{revtex4}

\usepackage{amssymb}
\usepackage{graphicx}
\usepackage{dcolumn}
\usepackage{bm}
\usepackage{color}

\begin{document}

\title{Coherent optical control of the spin of a single hole in a quantum dot}

\author{T.~M.~Godden}
\affiliation{Department of Physics
and Astronomy, University of Sheffield, Sheffield, S3 7RH, United
Kingdom}

\author{J.~H.~Quilter}
\affiliation{Department of Physics
and Astronomy, University of Sheffield, Sheffield, S3 7RH, United
Kingdom}

\author{A.~J.~Ramsay}
\email{a.j.ramsay@shef.ac.uk}
\affiliation{Department of Physics
and Astronomy, University of Sheffield, Sheffield, S3 7RH, United
Kingdom}

\author{Yanwen Wu}
\affiliation{Cavendish Laboratory, University of Cambridge, Cambridge, CB3 OHE, United
Kingdom}

\author{P.~Brereton}
\affiliation{Cavendish Laboratory, University of Cambridge, Cambridge, CB3 OHE, United
Kingdom}

\author{S.~J.~Boyle}
\affiliation{Department of Physics
and Astronomy, University of Sheffield, Sheffield, S3 7RH, United
Kingdom}

\author{I.~J.~Luxmoore}
\affiliation{Department of Physics
and Astronomy, University of Sheffield, Sheffield, S3 7RH, United
Kingdom}

\author{J.~Puebla-Nunez}
\affiliation{Department of Physics
and Astronomy, University of Sheffield, Sheffield, S3 7RH, United
Kingdom}


\author{A.~M.~Fox}
\affiliation{Department of Physics and Astronomy, University of
Sheffield, Sheffield, S3 7RH, United Kingdom}

\author{M.~S.~Skolnick}
\affiliation{Department of Physics and Astronomy, University of
Sheffield, Sheffield, S3 7RH, United Kingdom}

\date{\today}

\begin{abstract}
We demonstrate coherent optical control of a single hole spin confined to an InAs/GaAs quantum dot. A superposition of hole spin states is created by fast (10-100~ps) dissociation of a spin-polarized electron-hole pair. Full control of the hole-spin is achieved by combining coherent rotations about two axes: Larmor precession of the hole-spin about an external Voigt geometry magnetic field, and rotation about the optical-axis due to the geometric phase shift induced by a picosecond laser pulse resonant with the hole-trion transition.
\end{abstract}
\maketitle

The principal source of dephasing of an electron spin trapped on a quantum dot is the nuclear spins of the crystal-lattice \cite{Greilich_sci}. Since the heavy-hole has a p-type, rather than s-type wavefunction, the hyperfine interaction experienced by the hole is about one tenth of that of the electron due to the suppression of the contact term \cite{Fischer_prb,Fallahi_prl,Chekovich_prl}. This has stimulated interest in using the hole spin as a qubit, encouraged by measurements of ms-scale lifetimes \cite{Heiss_prb} and high visibility dips in coherence population trapping (CPT) experiments suggesting coherence times in the microsecond regime \cite{Brunner_sci}. Key requirements for the qubit are the ability to prepare, detect \cite{Gerardot_nat,Ramsay_prl,Godden_apl} and rotate a single hole spin. However, whilst the coherent optical control of a single electron spin is relatively advanced \cite{Press_nat}, there are no reports of the control of a hole spin.

Here we report the full coherent optical control of a single heavy-hole spin, $m_J=\pm 3/2$, confined to an InAs/GaAs quantum dot in an in-plane magnetic field. A coherent superposition of the energy-eigenstates of the hole spin is created through the ionization of a spin-polarized electron-hole pair, where the electron tunnels from the dot to leave a spin-polarized hole \cite{Ramsay_prl}, which then precesses about the applied magnetic field along the x-axis. From the decay of the hole spin precession,  a dephasing time $T_2^*=15.4_{-3.2}^{+5.5}~\mathrm{ns}$ is deduced. This value is consistent with dephasing due to fluctuations in a nuclear magnetic field acting on the hole spin, and is 7-13 times longer than for an electron spin confined to an InAs/GaAs quantum dot \cite{Press_nphoton}, as expected from the weaker hyperfine interaction. Rotation of the hole-spin about the optical z-axis is achieved using a $2\pi$ circularly polarized laser pulse resonant with the hole-trion transition to impart a geometric phase-shift on the selected spin. In this way we demonstrate the ability to perform any arbitrary rotation of the hole spin by combining rotations about two axes.

\begin{figure*}
\begin{center}
\includegraphics[scale=0.8,angle=0]{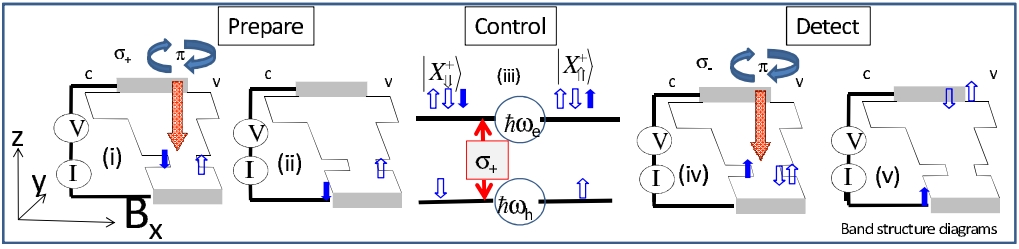}
\end{center}
\caption{Preparation, coherent control and detection of a single hole spin. (i) Resonant excitation of the neutral exciton transition by a laser pulse propagating along the z-axis creates a spin-polarized electron-hole pair.
(ii) When the electron tunnels it leaves a spin-polarized hole that precesses about the magnetic field applied along the x-axis. (iii) Rotation of hole-spin. The hole (trion) spin-z states are coupled with in-plane Zeeman energies of $\hbar\omega_h$ ($\hbar\omega_e$) respectively. The $\sigma_+$-polarized control pulse couples the $\mid\Downarrow\rangle\leftrightarrow\mid\Downarrow\Uparrow\downarrow\rangle$ states only, imparting a phase-shift on $\mid\Downarrow\rangle$. (iv) To detect the hole-spin, a circularly polarized laser pulse resonant with the hole-trion transition is absorbed conditional on the spin-z state of the hole. (v) When the additional carriers created in step (iv) tunnel from the dot a change in photocurrent proportional to the occupation of the hole spin state selected by the helicity of the detection pulse is measured.
}\label{fig:energy}
\end{figure*}

\begin{figure*}
\begin{center}
\includegraphics[scale=1.8]{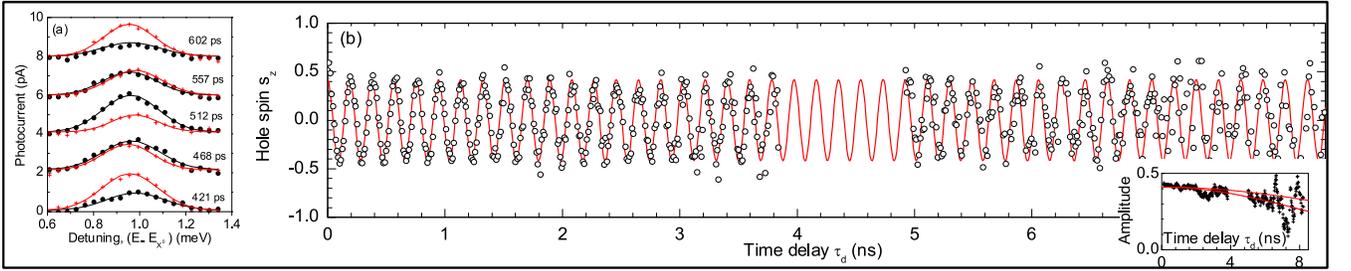}
\end{center}
\caption{ Precession of single hole spin ($\mathrm{B_x=4.7~T, V_{g}=0.8~V}$). (a) Change in photocurrent vs detection pulse detuning for co ($\bullet$) and cross ($+$) circular excitation at various time-delays. The peak corresponds to the hole-trion transition.  (b) Precession of hole spin, $s_z=\frac{I_{-}^{pc}-I_{+}^{pc}}{I_{-}^{pc}+I_{+}^{pc}}$ vs detection-pulse time-delay $\tau_d$. $I_{\pm}^{pc}$ is the amplitude of the photocurrent peaks measured for $\sigma_{\pm}$ polarized detection pulse as in (a).  (solid-line) undamped cosine to guide the eye. (inset) Amplitude of Larmor precession vs $\tau_d$, the traces are Gaussian decays with $T_L=12.2,~20.9~\mathrm{ns}$. The amplitude is determined from a sine fit to the data of (a) in the range $\tau_d\pm T/2$, where $T$ is the Larmor period.
}\label{fig:fig1}
\end{figure*}

The principle of the experiment is sketched in fig. \ref{fig:energy}. The InAs/GaAs quantum dot which is embedded in the intrinsic region of an n-i-Schottky diode structure. The sample is held at 4.2 K in a Helium bath cryostat, and a magnetic field is applied in-plane. A reverse bias is applied such that the electron tunneling rate is fast compared with the splitting between the energy-eigenstates of the neutral exciton spin states.  Due to a larger effective mass, the hole tunneling rate is much slower than for the electron. The sample is excited at normal incidence by two or three circularly polarized picosecond Gaussian laser pulses of 0.2-meV FWHM derived from a single 100-fs Ti:sapphire laser. A photocurrent detection technique is used \cite{Zrenner_nat}. A background photocurrent is subtracted from all data. For more details on the sample and the preparation of the laser pulses, see ref. \cite{supplement}.

The precession of a single hole spin in an applied magnetic field of 4.7~T, is observed by exciting the dot with two laser pulses termed preparation and detection, separated by a time-delay $\tau_d$. In step (i) of fig. \ref{fig:energy}, the $\sigma_+$ circularly polarized preparation pulse is tuned on resonance with the bright neutral exciton transition, and has a pulse-area of $\pi$. This creates a spin-polarized electron-hole pair $\mid\downarrow\Uparrow\rangle$.  This is a superposition of the linearly polarized eigenstates of fine-structure splitting $17~\mathrm{\mu eV}$, causing the exciton spin to precess. (ii) If the frequency mismatch between the exciton and hole spin precessions is small compared with the  electron tunneling rate \cite{Ramsay_prl,Godden_apl}, when the electron tunnels from the dot it leaves a hole with a net spin-up at time zero \cite{Godden_note}. (iii) The energy-eigenstates of the hole-spin are aligned along the external magnetic field $B_x$ and the spin-up state is a superposition of these states. This  causes the hole-spin to precess about $B_x$ at the Larmor frequency of the in-plane hole Zeeman splitting. (iv) To detect the hole spin, the frequency of the circularly polarized detection pulse, also of pulse-area $\pi$, is scanned through the hole-trion transition \cite{supplement} and a change in photocurrent recorded. Due to Pauli blockade, creation of two holes of the same spin is forbidden. (v) Therefore absorption of the detection pulse results in a change in photocurrent proportional to the occupation of the hole spin up/down state as selected by the helicity of the detection pulse. Examples of such two-color photocurrent spectra for co- and cross-circular excitation are presented in fig. \ref{fig:fig1}(a) as a function of the inter-pulse time-delay $\tau_d$. The amplitude of the peaks oscillate in anti-phase due to Larmor precession of the hole spin. The energy separation of the peaks also oscillates with $\tau_d$. This is probably a result of optical pumping of the nuclear spins, but lies outside the scope of this letter.

\begin{figure*}
\begin{center}
\includegraphics[scale=1.7]{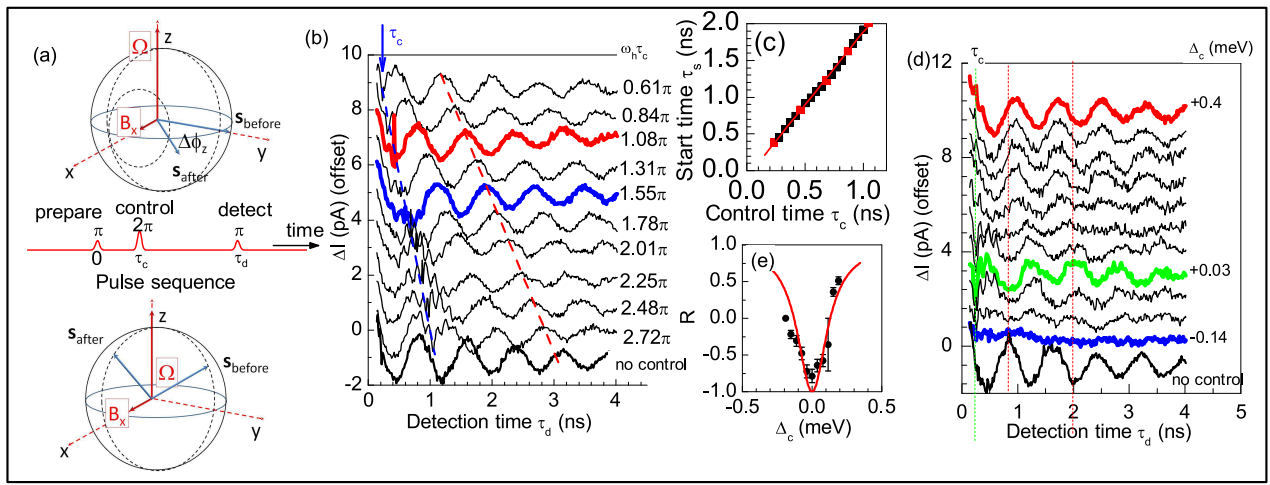}
\end{center}
\caption{Coherent control of hole spin. ($B=1.13~\mathrm{T}$, $V=0.8~\mathrm{V}$) (a) (Top,bottom) Orbits of Larmor precession of hole-spin about magnetic field axis x, before and after the control pulse are shown as dashed lines.  Top-sphere illustrates the experiments in (b), where an on-resonance control pulse rotates the hole-spin about z by angle $\pi$, changing phase of precession.  The bottom-sphere illustrates the experiments in (d). The control acts when the hole-spin points along y, rotating the hole-spin by $\Delta\phi_z(\Delta_c)$ about z, reducing the amplitude of precession.  (a,middle) Pulse sequence. (b) Control of phase of precession. $\Delta I=I_{-}^{pc}-I_{+}^{pc}$ is plotted vs detection-time $\tau_d$ for various control-times $\tau_c$.  (c) Change in start-time of the precession due to the control pulse: $\tau_s=1.99\tau_c-69~\mathrm{ps}$. (d) Control of rotation angle $\Delta\phi_z$ via the detuning $\Delta_c$ varies the amplitude of the  hole-spin precession. The control-time $\tau_c=234~\mathrm{ps}$, where the hole-spin points along y. (e)  ($\bullet$) Ratio $R$ of precession amplitude normalized to total hole population, with and without  the control vs $\Delta_c$. (line) Calculation of $\cos{\Delta\phi_z}$, the ideal dependence of  $R$ \cite{Economou_prl}.
}\label{fig:fig3}
\end{figure*}

Figure \ref{fig:fig1}(b) shows the precession of the hole-spin for a time-delay up to 8.5~ns. The z-component of the spin is calculated using $s_z=\frac{I^{pc}_--I^{pc}_+}{I^{pc}_-+I^{pc}_+}$, where $I^{pc}_{\pm}$ is the amplitude of the hole-trion peak measured for a detection pulse of $\sigma_{\pm}$ polarization, and plotted against the time-delay $\tau_d$. The frequency of the oscillation is proportional to the magnetic field, confirming that the oscillation arises from a coherent superposition of two Zeeman-split hole spin states with an in-plane hole g-factor of $g_{hx}=0.079\pm 0.004$. For the 0.8-V gate voltage used, the hole tunneling time is 4~ns. This is small compared to the 13-ns repetition period of the laser, ensuring the dot is empty on the arrival of the next preparation pulse, but long enough to enable over 40 periods of the precession to be  resolved. Due to hole tunneling, the total photocurrent signal of the trion peak becomes weak at large time-delays, leading to the increase in the scatter of the data.

By factoring out the hole tunneling, the damping of the Larmor precession in fig. \ref{fig:fig1}(b) depends on the relaxation and dephasing of the hole spin only. This assumes that the hole tunneling rate is independent of spin.
Since no spin-echo techniques are employed, the most likely source of hole-spin decoherence is dephasing due to inhomogeneous broadening. From Gaussian fits to the amplitude of the precession, shown in the inset of fig. \ref{fig:fig1}(b), where $s_z\propto \exp{(-\tau_d^2/T_L^2)}$, a damping time of $T_L=15.4_{-3.2}^{+5.5}~\mathrm{ns}$ is deduced. This is similar to the hole-spin dephasing time $T_2^*$ measured for an ensemble of InAs/GaAs dots \cite{Eble_prl}. It is 7-13 times longer than the 1.7~ns measured by Press {\it et~al} \cite{Press_nphoton} for an electron spin confined to a single  InGaAs/GaAs quantum dot. This is in-line with the ratio of the hyperfine interaction strengths of the electron and hole measured  for InAs/GaAs quantum dots \cite{Fallahi_prl}. Therefore we cautiously suggest that the main source of dephasing is the hole-nuclear spin interaction. To support this viewpoint, estimates of $T_{2}^*(GaAs)\approx 13~\mathrm{ns}$, and $T_{2}^*(InAs)\approx 5.4~\mathrm{ns}$  were calculated \cite{supplement}, in good semi-quantitative agreement with the measured $T_L$. The $T_L$ is small compared to the microsecond-scale dephasing time measured by Brunner {\it et~al} \cite{Brunner_sci} in a coherence population trapping (CPT) experiment. In the CPT experiments, the hole-spin is aligned along the in-plane magnetic field (x), whereas in our experiments, the hole-spin precesses in the yz-plane. We speculate that the anisotropy \cite{Fischer_prb} of the hole-hyperfine coupling leads to the differences in the measured $T_2^*$. The overall coherence time is limited by hole-tunneling and the repetition rate of the laser, but this could be overcome through dynamic control of the tunneling rates as in the experiments of ref. \cite{Heiss_prb}. We note that, although the $T_L$ measured here is large compared to an electron-spin in an InAs/GaAs quantum dot \cite{Press_nphoton}, it is similar to electron-spin values measured for much larger GaAs interface \cite{Mikkelsen_nphys,Dutt_prl} or electrically defined \cite{Koppens_prl} quantum dots, where longer dephasing times are to be expected, since the variance of the Overhauser field scales with the number of nuclei, $N$,  as $\sim N^{-1/2}$. If the carrier wavefunction of our dot is approximated as $\mid\psi\mid^2\sim e^{-r^2/a^2}$, then $a=3.2-3.5~\mathrm{nm}$, as deduced from measurements of exciton Rabi rotations \cite{Ramsay_prl2010}.

We now present experiments to demonstrate an arbitrary rotation of the hole spin about a second axis  using a third laser pulse termed the control pulse.
We use a `geometric-phase' approach as proposed theoretically in ref. \cite{Economou_prl} and demonstrated for an ensemble of electron spins in ref. \cite{Greilich_nphys}. If the hole-spin is represented by a vector on a Bloch-sphere, as depicted in fig. \ref{fig:fig3}(a),  the magnetic field leads to spin precession about the x-axis, and the control pulse to rotation about  the beam-path of the laser, ie. the z-axis. The control pulse has circular polarization and is resonant with the hole-trion transition, as shown in fig. \ref{fig:energy}(iii). On the timescale of the control pulse, the precessions of the hole and trion states are effectively stationary and the $\sigma_+$ polarized laser couples the $\mid\Downarrow\rangle \leftrightarrow \mid\downarrow\Uparrow\Downarrow\rangle$ states only. Initially, the hole spin is in a superposition state $\vert \psi\rangle=h_{\Uparrow}\mid\Uparrow\rangle+h_{\Downarrow}\mid\Downarrow\rangle$. The control pulse drives a  Rabi rotation between the selected hole spin and its corresponding trion state such that $\mid \psi\rangle \rightarrow h_{\Uparrow}\mid\Uparrow\rangle+h_{\Downarrow}[\cos{\Theta/2}\mid\Downarrow\rangle+i\sin{\Theta/2}\mid\downarrow\Uparrow\Downarrow\rangle]$, where $\Theta$ is the pulse-area.
In the ideal case of weak trion dephasing, and $\Theta=2\pi$, the state of the dot is returned to the hole-spin subspace having acquired a phase-shift of $\pi$ \cite{Economou_prl}. This is also true for detuned control pulses with a hyperbolic secant shape, similar to the Gaussian shape used here, except that the z-axis rotation-angle $\Delta\phi_z$ depends on the detuning  \cite{Economou_prl}.

We first present experiments demonstrating control of the phase of the hole-spin precession using a $2\pi$ control pulse. The magnetic field is reduced to 1.128~T, where the hole and trion Zeeman splittings of $5.1$ and $30~\mathrm{\mu eV}$ respectively are small compared to the bandwidth of the control pulse. For reference, the hole-spin precession with a period of 770~ps is measured without the control pulse and is shown as the lowest plot in figs. \ref{fig:fig3}(b,d). The detection pulse is resonant with the hole-trion transition and the difference between the photocurrents measured for  $\sigma_{\pm}$ detection pulses is plotted: $\Delta I=I^{pc}_--I^{pc}_+$. The $2\pi$-control pulse is tuned on resonance with the hole-trion transition and arrives at a time-delay of $\tau_c$ after the preparation pulse. The hole-spin precession is measured by scanning the detection time $\tau_d$, and a series of measurements for different values of $\tau_c$ are presented in fig. \ref{fig:fig3}(b). The main effect of the control pulse is to change the phase of the hole-spin precession as seen in fig. \ref{fig:fig3}(b). For detection times within plus or minus the electron-tunneling time, a fast 138-ps period oscillation is also observed. This is due to precession of a trion component created by the control pulse due to  the imperfect contrast of the hole-trion Rabi rotation \cite{Ramsay_prl2010}.

The red-trace in fig. \ref{fig:fig3}(b) presents the case where the hole-spin points along the z-axis when the control pulse arrives. Consequently, a rotation about the z-axis has minimal effect on the hole-spin as seen by comparing the bold and red-traces of fig. \ref{fig:fig3}(b). For the blue-trace, just before applying the control pulse, the hole-spin points along the y-axis and a rotation of $\pi$ about the z-axis phase-shifts the hole-spin precession by $\pi$. More generally, the effect of the  rotation is to reflect the hole-spin about the z-axis. The hole-spin before applying the control pulse can be written as  $\mathbf{s}=s^{(0)}(0, \sin{\omega_h\tau_c},\cos{\omega_h\tau_c})$. A reflection about the z-axis maps $\mathbf{s}\rightarrow s^{(0)}(0, \cos{\omega_h\tau_c}, \sin{\omega_h\tau_c})$, and subsequently the measured hole-spin precession evolves as $s_z=\cos{(\omega_h(\tau_d-2\tau_c))}$. In other words, the phase of the hole-spin is shifted by $-2\omega_h\tau_c$,  as occurs in a spin-echo experiment. The expected gradient of 2 for the phase of the hole-spin precession $\omega_h\tau_s$  is confirmed in fig. \ref{fig:fig3}(c), where $\tau_s$, defined with respect to the case of no control pulse, is found by fitting the time-traces of fig. \ref{fig:fig3}(b) to $\Delta I(\tau_d)=\Delta I^{(c)}\cos{(\omega_h(\tau_d-\tau_s))}$, for $\tau_d\gtrsim \tau_c+200~\mathrm{ps}$.

In the final set of experiments, we demonstrate control of the rotation angle $\Delta\phi_z$ induced by the control pulse via the detuning $\Delta_c$. The time-delay of the control pulse is set to $\tau_c=234~\mathrm{ps}$. On arrival of the control pulse, the hole-spin points along the y-axis, where $s_z$ is most sensitive to rotations about the z-axis. A series of hole-spin precessions are measured for different detunings $\Delta_c$ of the control pulse, and the results are presented in fig. \ref{fig:fig3}(d). The red-trace shows the case where the control pulse is far detuned from the hole-trion transition. The precession is relatively unaffected by the control, since the far-detuned pulse only induces a small rotation angle. As the control is tuned into resonance, the amplitude of the precession decreases. For a detuning of $-0.14~\mathrm{meV}$, which is approximately equal to the bandwidth  of the control pulse, the rotation angle $\Delta\phi_z$ is close to $\pi/2$. This leaves the hole-spin aligned along the x-axis which suppresses the subsequent precession of the hole-spin about the magnetic field as shown in the blue-trace. Near resonance, the amplitude changes sign indicating a rotation angle of greater than $\pi/2$. The amplitude of the hole-spin precession is maximal when the control is very close to resonance, as shown in green.

Figure \ref{fig:fig3}(e) is a plot of the ratio of the precession amplitudes, normalized to the total hole population, with and without the control pulse $R=s_z^{(c)}/s_z^{no}$ against the detuning of the control pulse $\Delta_c$. This is measured using a series of two-color photocurrent spectra as in fig. \ref{fig:fig1}(b). The red-line in fig. \ref{fig:fig3}(e) is a calculation of $R$ expected for the ideal case of no trion dephasing, namely $R=\cos{(\Delta\phi_z)}$, where $\tan{(\Delta\phi_z/2)}=\Delta\omega_c/\Delta_c$, with a bandwidth $\Delta\omega_c=0.13~\mathrm{meV}$ \cite{Economou_prl}. There is close agreement between experiment and theory, which implies that the control-pulse rotates the hole-spin by a detuning-dependent angle $\Delta\phi_z$, with a maximum value close to $\pi$, in accordance with model of ref. \cite{Economou_prl}.

In conclusion, by combining coherent rotations about two axes,  defined by an external magnetic field and the optical axis of a control laser, full control of the hole-spin on the Bloch-sphere is achieved.   The optical rotation has a gate-time defined by the 14~ps FWHM of the control pulse, which is much shorter than the measured extrinsic dephasing time of the hole spin $T_L=15.4_{-3.3}^{+5.5}~\mathrm{ns}$.

We thank the EPSRC (UK) EP/G001642, and the QIPIRC UK  for financial support; H.~Y.~Liu and M.~ Hopkinson for sample growth; O.~Tsyplyatyev, E.~A.~Chekhovich and J.~J.~Finley for discussions.

On completion of the experiments we became aware of related experiments, see ref. \cite{Greilich_hole,DeGreve_ArXiv}.


\end{document}